\documentclass[twocolumn,aps,prl,superscriptaddress,showpacs]{revtex4}

\input epsf 
\def \roots {\sqrt{s_{\rm NN}}} 
\def \pbarp {\overline{p}/{p}} 
\def \pt    {p_\perp} 
\def \mpt    {\langle p_\perp \rangle} 
\def \mt    {m_\perp} 
 
\def \dndy  {dN/dy} 
     
\def \kpi {K^{-}/\pi^{-}}

\def \pbarpi {\bar{p}/\pi^{-}}
\begin{document}

\title{
\vspace{0.2in}
Identified particle distributions in $pp$ and Au+Au collisions at $\roots=200$~GeV}

%

\affiliation{Argonne National Laboratory, Argonne, Illinois 60439}
\affiliation{Brookhaven National Laboratory, Upton, New York 11973}
\affiliation{University of Birmingham, Birmingham, United Kingdom}
\affiliation{University of California, Berkeley, California 94720}
\affiliation{University of California, Davis, California 95616}
\affiliation{University of California, Los Angeles, California 90095}
\affiliation{Carnegie Mellon University, Pittsburgh, Pennsylvania 15213}
\affiliation{Creighton University, Omaha, Nebraska 68178}
\affiliation{Nuclear Physics Institute AS CR, \v{R}e\v{z}/Prague, Czech Republic}
\affiliation{Laboratory for High Energy (JINR), Dubna, Russia}
\affiliation{Particle Physics Laboratory (JINR), Dubna, Russia}
\affiliation{University of Frankfurt, Frankfurt, Germany}
\affiliation{Indiana University, Bloomington, Indiana 47408}
\affiliation{Insitute  of Physics, Bhubaneswar 751005, India}
\affiliation{Institut de Recherches Subatomiques, Strasbourg, France}
\affiliation{University of Jammu, Jammu 180001, India}
\affiliation{Kent State University, Kent, Ohio 44242}
\affiliation{Lawrence Berkeley National Laboratory, Berkeley, California 94720}\affiliation{Max-Planck-Institut f\"ur Physik, Munich, Germany}
\affiliation{Michigan State University, East Lansing, Michigan 48824}
\affiliation{Moscow Engineering Physics Institute, Moscow Russia}
\affiliation{City College of New York, New York City, New York 10031}
\affiliation{NIKHEF, Amsterdam, The Netherlands}
\affiliation{Ohio State University, Columbus, Ohio 43210}
\affiliation{Panjab University, Chandigarh 160014, India}
\affiliation{Pennsylvania State University, University Park, Pennsylvania 16802}
\affiliation{Institute of High Energy Physics, Protvino, Russia}
\affiliation{Purdue University, West Lafayette, Indiana 47907}
\affiliation{University of Rajasthan, Jaipur 302004, India}
\affiliation{Rice University, Houston, Texas 77251}
\affiliation{Universidade de Sao Paulo, Sao Paulo, Brazil}
\affiliation{University of Science \& Technology of China, Anhui 230027, China}
\affiliation{Shanghai Institute of Nuclear Research, Shanghai 201800, P.R. China}
\affiliation{SUBATECH, Nantes, France}
\affiliation{Texas A\&M, College Station, Texas 77843}
\affiliation{University of Texas, Austin, Texas 78712}
\affiliation{Valparaiso University, Valparaiso, Indiana 46383}
\affiliation{Variable Energy Cyclotron Centre, Kolkata 700064, India}
\affiliation{Warsaw University of Technology, Warsaw, Poland}
\affiliation{University of Washington, Seattle, Washington 98195}
\affiliation{Wayne State University, Detroit, Michigan 48201}
\affiliation{Institute of Particle Physics, CCNU (HZNU), Wuhan, 430079 China}
\affiliation{Yale University, New Haven, Connecticut 06520}
\affiliation{University of Zagreb, Zagreb, HR-10002, Croatia}
\author{J.~Adams}\affiliation{University of Birmingham, Birmingham, United Kingdom}
\author{C.~Adler}\affiliation{University of Frankfurt, Frankfurt, Germany}
\author{M.M.~Aggarwal}\affiliation{Panjab University, Chandigarh 160014, India}
\author{Z.~Ahammed}\affiliation{Variable Energy Cyclotron Centre, Kolkata 700064, India}
\author{J.~Amonett}\affiliation{Kent State University, Kent, Ohio 44242}
\author{B.D.~Anderson}\affiliation{Kent State University, Kent, Ohio 44242}
\author{M.~Anderson}\affiliation{University of California, Davis, California 95616}
\author{D.~Arkhipkin}\affiliation{Particle Physics Laboratory (JINR), Dubna, Russia}
\author{G.S.~Averichev}\affiliation{Laboratory for High Energy (JINR), Dubna, Russia}
\author{S.K.~Badyal}\affiliation{University of Jammu, Jammu 180001, India}
% ...

\collaboration{STAR Collaboration}\homepage{www.star.bnl.gov}\noaffiliation


\begin{abstract}
Transverse mass and rapidity distributions for charged pions, charged kaons, protons and antiprotons are reported for  $\roots = 200$~GeV $pp$ and Au+Au collisions at RHIC.  
The transverse mass distributions are rapidity independent within $|y|<0.5$, consistent with a boost-invariant system in this rapidity interval.
Spectral shapes and relative 
particle yields are similar in $pp$ and peripheral Au+Au collisions and change smoothly to central Au+Au collisions.
No centrality dependence was observed in the kaon and antiproton production rates relative to the pion
production rate from medium-central to central collisions.
Chemical and kinetic equilibrium model fits to our data reveal strong radial flow and relatively long duration from chemical to kinetic freeze-out in central Au+Au collisions.  The chemical freeze-out temperature appears to be independent of initial conditions at RHIC energies. 

\end{abstract}


\maketitle

Quantum chromodynamics predicts the existence of a new form of matter, the quark-gluon plasma (QGP), at extreme conditions of high energy density, possibly achieved in relativistic heavy ion collisions \cite{karsch}.
Signals of QGP may remain in the bulk properties of the collision, and simultaneous observations of multiple QGP signals in the final state would serve as a strong evidence of QGP formation. 
These bulk properties include strangeness and baryon production rates 
and collective transverse radial flow. These can be studied via particle spectra.

In this letter we report results on charged pion ($\pi^{\pm}$), charged kaon ($K^{\pm}$), proton ($p$) and antiproton ($\bar{p}$) production from  $pp$ and Au+Au collisions at RHIC by the STAR experiment at the nucleon-nucleon center-of-mass energy of $\roots = 200$~GeV. 
In some models it is argued that particle multiplicity density per transverse area of interaction measures the initial gluon density~\cite{zxu2}, particle ratios measure the chemical freeze-out conditions~\cite{new2}, and transverse momentum spectra measure the kinetic freeze-out conditions~\cite{BW}. We study these properties at mid-rapidity as a function of centrality. 
The rapidity dependences of particle production and 
spectra shape are also investigated. 

Charged particles are detected in the STAR Time Projection Chamber (TPC)~\cite{starNIM}. The TPC is surrounded by  a solenoidal magnet providing a uniform magnetic field of 0.5~T along the beam line.  Zero degree calorimeters
and beam-beam counters~\cite{highpt} provide a minimum bias trigger for Au+Au and $pp$ collisions, respectively.
Events with a primary vertex within $\pm25$~cm of the geometric center of the
TPC along the beam axis are accepted.
For this analysis, about $2.0\times 10^6$ Au+Au and about $2.5 \times 10^6$ $pp$  minimum bias accepted events are used.  
Only primary tracks - tracks pointing to the primary vertex within 3~cm - are selected. 
The Au+Au events are divided into 9 centrality classes based on measured charged particle multiplicity within pseudo-rapidity  $| \eta | < 0.5$. 
These classes consist, from central to peripheral, of 0-5\%, 5-10\%, 10-20\%, 20-30\%, 30-40\%, 40-50\%, 50-60\%, 60-70\%, and 70-80\% of the 
geometrical cross-section.

Particle identification is accomplished by measuring the ionization energy loss $dE/dx$. 
The mean $\langle dE/dx \rangle$ is determined from 70\% of the samples with the lowest $dE/dx$ along a track. 
To insure good momentum and $\langle dE/dx \rangle$ resolution tracks are required to have at least 25 out of the maximum 45 hits in the TPC.  
 The $\langle dE/dx \rangle$ resolution varies between 6\% and 10\% from $pp$ to central Au+Au events. 
The reconstructed  momenta  are corrected for most likely energy loss in the detector. 
The correction is negligible for $\pi^{\pm}$, under 2\% for $K^{\pm}$ and under 5\% for $p$ and $\bar{p}$ in the covered momentum ranges.
The momentum resolution was estimated  to be about $2\%$ at $\pt$ = 0.5~GeV/c. 
Uncorrected particle yields are extracted from  $\langle dE/dx \rangle$ distributions for each $\pt$, rapidity  and centrality bin~\cite{pbar,p,pi_star,130kaon}.

Corrections are applied to account for  tracking inefficiency, detector acceptance, hadronic interactions, and particle decays.  
The
total reconstruction efficiencies are obtained from embedding Monte Carlo (MC) tracks into real events at the raw data level and subsequently reconstructing these events.  
The propagation of single tracks is calculated using  GEANT, a detailed  description of the STAR geometry, and a realistic simulation of the TPC response \cite{pbar,p,pi_star,130kaon}.   
The efficiencies 
for $\pi^{\pm}$ are 50-70\% 
and 80-90\% 
in the covered $\pt$ for the 0-5\% and 70-80\% events, respectively. The corresponding efficiencies for $K^{\pm}$ are  40-70\% and 20-50\% and for $p$ and $\bar{p}$ 70-75\% and 75-80\%. 
Background protons knocked out from the detector material are subtracted. 
This background 
is 50-60\% at $\pt = 0.4$~GeV/$c$ and becomes less than 5\% at $1.0$~GeV/$c$ \cite{p}.

Corrections for the $pp$ data  are similar to those for the 70-80\%  Au+Au events. 
Additional corrections are applied  for primary vertex reconstruction inefficiency and fake events (events with mis-reconstructed vertex due to pile-up background). 
These corrections are obtained by embedding HIJING~\cite{hijing} events into events that had been triggered on empty bunches, and reconstructing  the combined events.
The vertex reconstruction inefficiency strongly decreases with increasing event multiplicity resulting in approximately 14\% of events being missed, over 80\%  of which have fewer than three tracks  in the TPC. 
About 12\% of $pp$ events are fake events with
reconstructed multiplicity 
about half of that of real events 
due to time distortion in the pile-up background, 
resulting in an overall correction of 6-8\% 
in the covered $\pt$ range.

The pion spectra are further corrected for weak decay products, muon contamination and background pions produced in the detector material. The weak decay correction is approximately 12\% and was estimated from the measured $K^0_s$ and $\Lambda$ distributions~\cite{lambda,130kaon} extrapolated to our energy. 
Because weak decay (anti)protons carry most of the parent momentum, their tracks behave as those originating from the primary vertex, resulting in the same reconstruction efficiency for weak decay and primary (anti)protons over the measured $\pt$ range.

The inclusive (anti)protons closely reflect total (anti)baryon production \cite{pbar,p}.  Therefore we present inclusive proton and antiproton distributions that are not corrected for weak decays. Based on the measured $\Lambda$ distribution~\cite{lambda}, we estimate that about 40\% of the measured protons are from weak decays, and the measured inclusive $\mpt$ are similar to those of primary protons.

The point to point systematic uncertainties on the spectra are estimated 
by varying event and track selection and analysis cuts and by assessing sample purity from the $dE/dx$ measurement. The estimated uncertainties are less than 4\% for $\pi^{\pm}$, $p$ and $\bar{p}$. 
Those for $K^{\pm}$ are less than 12\% for $\pt$ bins with significant overlap in $dE/dx$ with $e^{\pm}$ or $\pi^{\pm}$, and less than 4\% for other bins.   
An additional systematic error on the proton spectra due to background subtraction is estimated to be 5\% at low $\pt$ and negligible at high $\pt$ \cite{p}.
A correlated systematic uncertainty of 5\% is estimated for all spectra and is dominated by uncertainties in the MC determination of reconstruction efficiencies.

\begin{figure*}[bt]
\centerline{\epsfxsize=0.6\textwidth\epsfbox[100 20 510 330]{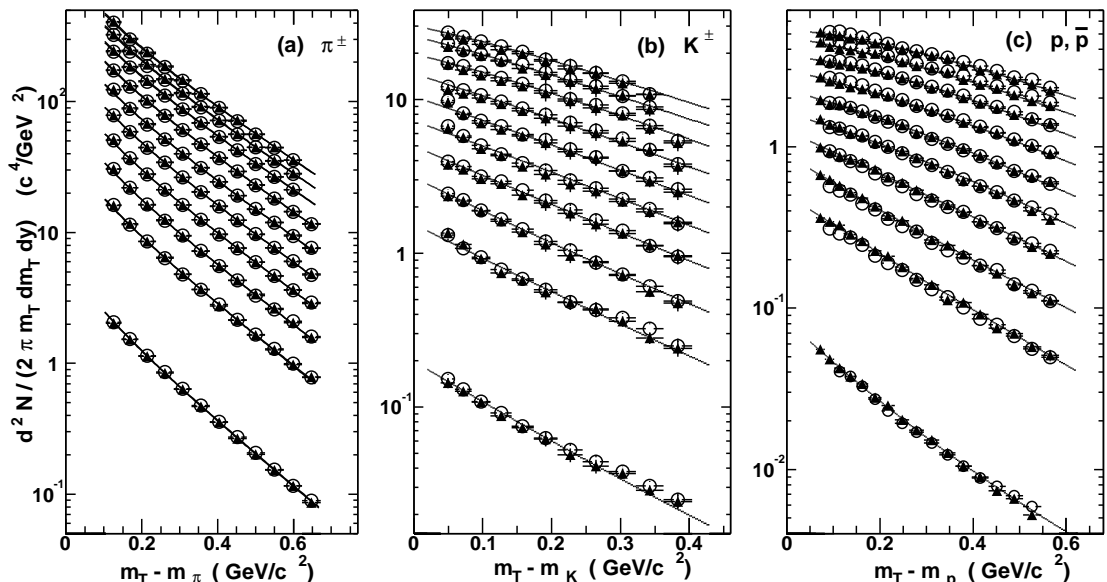}}
\caption{
Invariant yield as function of transverse mass for
$\pi^{\pm}$, $K^{\pm}$, and inclusive $p$ and  $\bar{p}$ 
at mid-rapidity ($|y|<0.1$) for $pp$ (bottom) and Au+Au events from 70-80\% (second bottom) to the 0-5\%
centrality bin (top).
Statistical and point-to-point systematic errors have been added in quadrature. 
Additional correlated systematic error due to uncertainty in the normalization is estimated to be 5\%.
Open circles	are for positive particles (all proton spectra are scaled  by 0.8), and closed triangles are for negative particles. The curves shown (Bose-Einstein fits for $\pi^-$ and blast-wave model fits for $K^-$ and $\bar{p}$) are explained in the text.
}
\label{spectra}
\end{figure*}

Figure 1 shows transverse mass ($\mt = \sqrt{\pt^2 + m^2}$) spectra  for $\pi^{\pm}$, $K^{\pm}$, $p$ and $\bar{p}$ for $pp$ and all centrality bins of Au+Au data within $|y| < 0.1$. For clarity, proton spectra are scaled by 0.8. 
Particle and anti-particle spectra shapes are similar for each centrality bin. 
While the $\pi^{\pm}$ spectra shapes are similar for $pp$ and Au+Au, $K^{\pm}$, $p$  and $\bar{p}$ spectra  show a progressive flattening  from $pp$ to central Au+Au events. Our $pp$ results are consistent with previous measurements at similar multiplicities \cite{E735}.  

The blast-wave model $-$ a hydrodynamically motivated  model with a kinetic freeze-out temperature $T_{kin}$ and a transverse flow velocity field $\beta$~\cite{BW} $-$  can simultaneously fit the $K^{\pm}$, $p$ and $\bar{p}$ spectra and the high-$\pt$ part ($\pt > 0.50$~GeV/{\it c}) of the $\pi^{\pm}$ spectra. 
We used a velocity profile of $\beta = \beta_{s} (r/R)^{n}$, where 
$r \leq R$ (the term $r/R$ accounts for the change in the velocity as a function of radial distance),
$\beta_{s}$ is the surface velocity, and $n$ is treated as a free parameter.
The value of  $n$ ranges from  $1.50 \pm 0.29$ in peripheral to $0.82 \pm 0.02$ in  central events. 
The fit results are superimposed in Fig. 1(b,c).  The obtained fit parameters for the 0-5\% Au+Au events are $T_{kin} = 89 \pm 10$~MeV and $\langle \beta \rangle = 0.59 \pm 0.05$, $\beta_s = 0.84 \pm 0.07$, and are similar to the 130 GeV results reported in~\cite{pi_star, flow}.
The systematic uncertainties in the fit parameters 
are estimated by excluding the kaon or the (anti)proton spectra from the fit.

Recent attempts to fit the measured RHIC spectra with a single (chemical and kinetic) freeze-out temperature claim 
this is possible 
if all the resonance and weak decay feed-downs are taken into account~\cite{bronowski}. Our MC study 
of that scenario shows significantly higher $\chi^{2}/$NDF 
compared to our blast-wave fits.

The low-$\pt$ part of the pion spectrum deviates from the blast-wave model description, possibly due to large contributions from resonances at low $\pt$. 
We fit the pion spectra to a
Bose-Einstein distribution ($\propto 1/(\exp{\frac{\mt}{T}}-1)$), the results of which are superimposed in Fig.1(a).
The yields outside the measured $\pt$ region are extrapolated using the blast-wave model for $K^{\pm}$, $p$ and $\bar{p}$ and the Bose-Einstein distribution for $\pi^{\pm}$. The uncertainties on these extrapolations are estimated by comparing to results using other functional forms. 
The estimated extrapolation uncertainties in the $\mpt$ and total yield are 5\% for $\pi^{\pm}$ and 5 to 10\% for $K^{\pm}$, $p$ and $\bar{p}$ (varying from $pp$ to central Au+Au collisions). 
For the 0-5\% Au+Au collisions, the integrated yields are $\dndy = 322 \pm 32$ for $\pi^+$,  $327 \pm 33$ for $\pi^{-}$, $51.3 \pm 7.7$ for $K^+$,  $49.5 \pm 7.4$ for $K^{-}$,  $34.7 \pm 6.2$ for $p$ and $26.7 \pm 4.0$ for $\bar{p}$.
The obtained $\pbarp$ ratio for the 0-5\% Au+Au collisions is $0.77 \pm 0.05$, indicating a nearly net-baryon free mid-rapidity region at this RHIC energy.

We extract the fiducial dN/dy by summing up the yields within the $\pt$ range of 0.20-0.70~GeV/{\it c} for $\pi^-$, 0.25-0.60~GeV/{\it c} for $K^-$, and  0.50-1.05~GeV/{\it c} for $\bar{p}$. 
Figure 2 depicts the rapidity dependence of the fiducial $\dndy$  
and extrapolated $\mpt$
for the 0-5\%  and 70-80\%  Au+Au events. 
We  do not observe changes in either shape or yield for any particle species within $|y|<0.5$. The $pp$ data and all other centrality bins of the Au+Au data exhibit the same behavior.
Such an absence of rapidity dependence of particle spectra was also observed for $\pi^{\pm}$, $p$ and $\bar{p}$ at $\roots = 130$~GeV Au+Au collisions \cite{p,pi_star}. This uniformity indicates the development of a boost-invariant region within the measured kinematic ranges. 

\begin{figure}[h]
\centerline{\epsfxsize=0.5\textwidth\epsfbox[ 0 0 567 301]{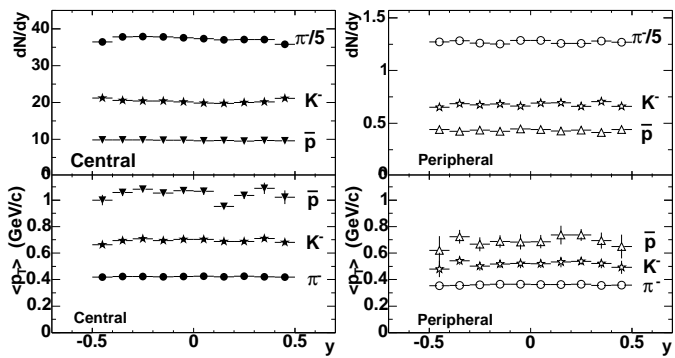}}
\caption{
Rapidity distributions of the fiducial 
yields  and integrated $\mpt$   for the 0-5\% and 70-80\%   Au+Au collisions. Pion yields are scaled down by a factor of 5. Errors shown are those propagated from Fig.1. Systematic errors on the fiducial yields are 5\%; those on $\mpt$
are 5\% for pions and  5-10\% for kaons and antiprotons.  
}
\label{dndy}
\end{figure}

The centrality dependence of the extracted $\mpt$ within $|y|<0.1$ is shown in Fig.~3(a). A smooth changeover from $pp$ to peripheral Au+Au collisions is observed for all particle species. 
The $\mpt$ increases from $pp$ and peripheral Au+Au to central Au+Au collisions, especially for $p$, $\bar{p}$ and $K^{\pm}$. This behavior is consistent with an increase of radial flow with collision centrality.

\begin{figure*}[hbt]
\centerline{\epsfxsize=0.95\textwidth\epsfbox[0 0 567 134]{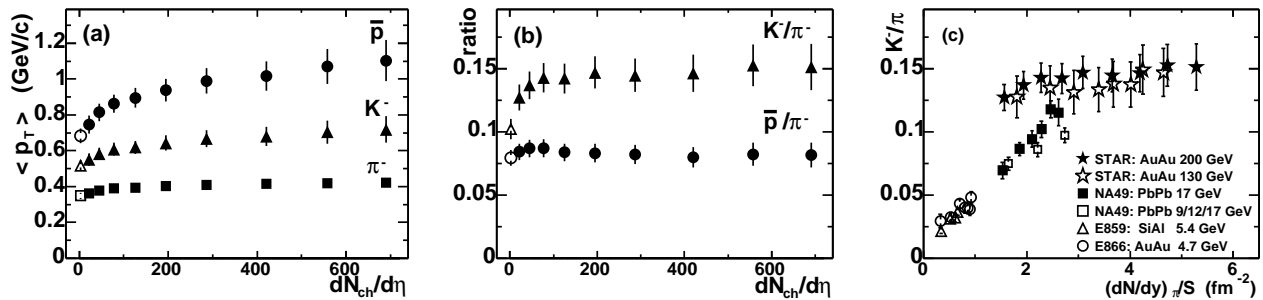}}
\caption{ 
(a) Mean transverse momentum of negative particles  and  (b) $\kpi$ and $\pbarpi$ ratios as function of the charged  hadron multiplicity. Open symbols are for $pp$, and filled ones are for Au+Au data.
(c) Mid-rapidity $K^{-} / \pi$ ratio as function  of $\frac{(\dndy)_{\pi}}{S}$. 
Systematic errors are shown for STAR data, and statistical errors for other data.
}
\label{ratio}
\end{figure*}

The $\kpi$ and $\pbarpi$ ratios of the integrated $\dndy$ yields within $|y|<0.1$ are depicted in Fig.~3(b).
We 
observe little centrality dependence of the $\kpi$ or $\pbarpi$ ratio from mid-central to central Au+Au collisions, indicating a similar 
freeze-out condition in these collisions. Similar centrality behavior has been observed for other particle ratios measured at $\roots = 200$~GeV and $130$~GeV \cite{130kaon,p}.

The observed centrality independence of  $\kpi$ is in  contrast to low energy data at SPS \cite{NA49} and AGS \cite{E802}, 
where a continuous increase in  $K^{-}/\pi$  was observed, roughly doubling from peripheral to central collisions.
To put our results into perspective with low energy data, we plot in Fig.~3(c) the $K^{-}/\pi$ ratio as a function of 
$\frac{(\dndy)_{\pi}}{S}$,
in an attempt to reflect effects of both  the collision energy and centrality. 
Here $S$ is an estimate of the transverse overlap area  based on the number of participants~\cite{pi_star}, experimentally measured for the AGS and SPS data and  calculated via the MC Glauber model for RHIC data~\cite{pi_star}.
The $\frac{(\dndy)_{\pi}}{S}$  may be related to the initial conditions of the collision~\cite{zxu2,Bjorken}, 
such as energy density. 
In high energy collisions the initial gluon density is saturated up to a momentum scale that is proportional to 
$\sqrt{ \frac{ (\dndy)_{\pi} }  {S} }$~\cite{zxu2}. 
Using data 
over a wide range of collision energy  measured in various colliding systems,
Fig.~3(c) shows a distinct change in the ratio behavior. Low energy measurements (each representing approximately top 60\% of the geometrical cross section) appear to follow a trend that saturates at RHIC energies.
One interpretation of this is that strangeness production at low energies depends on how the collision was initially  prepared, but  not at RHIC energies.
On the other hand, the $K^{+} / \pi$ and $\bar{p}/\pi$ ratios do not reveal a common trend with $\frac{(\dndy)_{\pi}}{S}$. However, we note that the net-baryon density, significant at low energies,  greatly affects $K^+$ and $\bar{p}$ abundances through associated production of $K^+$ with baryons \cite{130kaon} and baryon-antibaryon annihilation~\cite{ref3}, respectively.

In the framework of a chemical-equilibrium model~\cite{new2,Nu}, integrated yield ratios can be described by a set of parameters: the chemical freeze-out temperature ($T_{ch}$), the baryon and strangeness chemical potentials ($\mu_{B}$, $\mu_{s}$), and the strangeness suppression factor ($\gamma_{s}$). We fit our measured ratios with such a model to extract these parameters. 
The value obtained for the chemical potential, $\mu_{B} \approx 22 \pm 4$~MeV, is independent of 
centrality within errors, and $\mu_{s}$ is consistent with 0.
The obtained $\gamma_{s}$ increases from $0.56 \pm 0.04$ in $pp$ to $0.86 \pm 0.11$ in central Au+Au collisions reflecting the measured $K / \pi$ ratios. The obtained $T_{ch}$ is summarized in Fig.~4 as a function of charged hadron multiplicity, together with $T_{kin}$ and $\langle \beta \rangle$ extracted from the blast-wave model fit to our data.
As seen in Fig.~4,$\sqrt{\frac{(\dndy)_{\pi}}{S}}$ increases with centrality, $T_{ch}$ is independent of it, $T_{kin}$ decreases and $\langle \beta \rangle$ increases with centrality. 
This 
suggests that  Au+Au collisions  of different initial conditions always evolve 
to the same chemical freeze-out condition, and then cool down further to 
a kinetic freeze-out dependent on centrality. 
The expansion of the system gives rise to collective flow.

During  expansion from chemical to kinetic freeze-out,  entropy density drops approximately as $T^3$ \cite{ref2}, 
implying that the system size at  kinetic freeze-out is at least a factor of $\frac{T_{ch}}{T_{kin}}$ of the size at  chemical freeze-out. This suggests a time span from chemical to kinetic freeze-out in central collisions is at least of the order of 
$
\Delta t \approx (\frac{T_{ch}}{T_{kin}} - 1)R/ \beta_{s} \approx 6
$~fm/{\it c}.
Here we have taken $R = 6$~fm, the Au nuclei radius, as an estimate of the system size at chemical freeze-out.

\begin{figure}[hbt]
\centerline{\epsfxsize=0.4\textwidth\epsfbox[30 0 537 368]{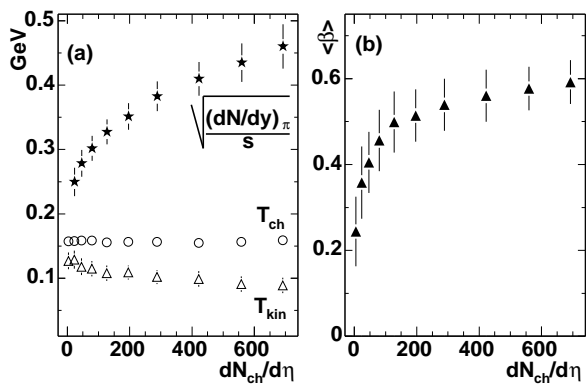}}
\caption{ (a) $\sqrt{\frac{(\dndy)_{\pi}}{S}}$
(stars), $T_{ch}$ 
(circles) and $T_{kin}$ 
(triangles) 
 and (b) $\langle \beta \rangle$ 
as a function of the charged  hadron multiplicity. Errors are systematic.}
\label{tfit}
\end{figure}

In summary, we have reported transverse mass and rapidity distributions of $\pi^{\pm}$, $K^{\pm}$, $p$ and $\bar{p}$ for $pp$ and Au+Au collisions at  $\roots = 200$~GeV at RHIC. A boost-invariant region of at least $\Delta y \approx 1$ is developed at mid-rapidity for particles within our measured $\pt$ range.  
The spectra are well described by the 
blast-wave model, yielding a decreasing $T_{kin}$ and increasing  $\langle \beta \rangle$ with centrality, reaching the values of  $T_{kin}= 89 \pm 10$~MeV and $\langle \beta \rangle = 0.59 \pm 0.05$ in the 5\% most central collisions. 
Particle ratios vary smoothly from $pp$ to peripheral Au+Au and remain relatively constant from mid-central to central Au+Au collisions. 
The $K^{-}/ \pi$ ratio from various collisions over a wide range of energy reveals a 
distinct behavior in $\frac{(\dndy)_{\pi}}{S}$.
A chemical equilibrium model fit to the  ratios yields a $T_{ch}$
insensitive to centrality with a value of $157 \pm 6$~MeV for the 5\% most central collisions. 
The drop in temperature from $T_{ch}$ to $T_{kin}$ and the development of strong radial flow suggest a significant expansion and long duration from chemical to kinetic freeze-out in central collisions.
From these results the following picture seems to emerge at RHIC: collision systems with varying initial conditions 
always evolve towards the same chemical freeze-out condition followed by cooling and expansion of increasing magnitude with centrality.

We thank the RHIC Operations Group and RCF at BNL, and the
NERSC Center at LBNL for their support. This work was supported
in part by the HENP Divisions of the Office of Science of the U.S.
DOE; the U.S. NSF; the BMBF of Germany; IN2P3, RA, RPL, and
EMN of France; EPSRC of the United Kingdom; FAPESP of Brazil;
the Russian Ministry of Science and Technology; the Ministry of
Education and the NNSFC of China; SFOM of the Czech Republic,
DAE, DST, and CSIR of the Government of India; the Swiss NSF.

 \end{document}